\documentclass{article}

\usepackage{arxiv}

\usepackage[utf8]{inputenc} 
\usepackage[T1]{fontenc}    
\usepackage{hyperref}       
\usepackage{url}            
\usepackage{booktabs}       
\usepackage{amsfonts}       
\usepackage{nicefrac}       
\usepackage{microtype}      
\usepackage{lipsum}
\usepackage{graphicx}
\graphicspath{ {./images/} }

\usepackage{lineno,hyperref}
\usepackage{amsmath,amsfonts,amssymb,mathtools}
\usepackage{siunitx}
\usepackage{subfigure}
\usepackage{caption}
\usepackage{flushend}
\usepackage{bm}
\usepackage[ruled,vlined]{algorithm2e}
\usepackage{algorithmic}

\title{Sparse regularization with a non-convex penalty for SAR imaging and autofocusing}

\author{
 Zi-Yao Zhang\thanks{This work was supported in part by a Chinese Scholarship Council PhD studentship (to Zhang) and in part by the Engineering and Physical Sciences Research Council (EPSRC) under grant EP/R009260/1 (AssenSAR).} \\
  Visual Information Laboratory\\
  University of Bristol\\
   \And
 Odysseas Pappas \\
  Visual Information Laboratory\\
  University of Bristol\\
  \And
 Igor G. Rizaev \\
  Visual Information Laboratory\\
  University of Bristol\\
  \And
 Alin Achim \\
  Visual Information Laboratory\\
  University of Bristol\\
}

\begin{document}
\maketitle

\begin{abstract}
In this paper, SAR image reconstruction with joint phase error estimation (autofocusing) is formulated as an inverse problem. An optimization model utilising a sparsity-enforcing Cauchy regularizer is proposed, and an alternating minimization framework is used to solve it, in which the desired image and the phase errors are optimized alternatively. For the image reconstruction sub-problem ($\bm{f}$-sub-problem), two methods are presented capable of handling the problem's complex nature, and we thus present two variants of our SAR image autofocusing algorithm. Firstly, we design a complex version of the forward-backward splitting algorithm (CFBA) to solve the $\bm{f}$-sub-problem iteratively. For the second variant, the Wirtinger alternating minimization autofocusing (WAMA) method is presented, in which techniques of Wirtinger calculus are utilized to minimize the complex-valued cost function in the $\bm{f}$-sub-problem in a direct fashion. For both methods, the phase error estimation sub-problem is solved by simply expanding and observing its cost function. Moreover, the convergence of both algorithms is discussed in detail. By conducting experiments on both simulated scenes and real SAR images, the proposed method is demonstrated to give impressive autofocusing results compared to other state of the art methods.
\end{abstract}


\section{Introduction}
%
%
%
%
Synthetic aperture radar (SAR) has become one of the most employed modalities in the field of remote sensing, largely due to its capability to collect data in all kinds of weather and lighting conditions. As a coherent imaging radar system often mounted on an airplane or satellite platform, it can transmit signals after frequency modulation to a certain scene on the ground and record the return radar echoes in flight. As with other coherent radar systems, these raw radar returns must then be processed to form an image suitable for visual interpretation. A detailed introduction on the working mechanisms of SAR can be found in \cite{moreira2013tutorial, ouchi2013recent}. 

The SAR data acquisition process is unfortunately frequently plagued by phase errors. Due to inaccuracies in the SAR platform trajectory measurement as well as possible existence of moving targets in the observed scene, the acquired data (radar returns) will contain phase errors. These phase errors in turn result in a defocusing effect in the formed SAR images. Techniques aiming at the direct estimation of these phase errors from the raw SAR data and the removal of them so as to improve the quality of the reconstructed SAR images are called autofocusing techniques.

Among the earliest autofocusing techniques, phase gradient autofocus (PGA) \cite{wahl1994phase} is a very well-known method. It first circularly shifts and windows the data, then uses these processed data to estimate the gradient of the phase errors, and finally integrates the estimations to obtain the phase errors themselves. Mapdrift autofocus is another classical technique for SAR autofocusing \cite{calloway1994subaperture}. It partitions the whole aperture into several sub-apertures from which the images (maps) are reconstructed, and measures the drift between each pair of maps to obtain the phase error. These classic methods often serve as the basis for newer methods, such as the variants of the mapdrift and phase gradient autofocus algorithms presented in \cite{tsakalides2001}, which employ fractional lower-order moments of the alpha-stable distribution for modelling the phase history data.

Differing from them, various methods based on optimization techniques have been proposed in recent years. Many of these methods belong to one of two categories. The first involves the construction of a sharpness metric to be maximized. For instance, a power function is chosen as the sharpness metric in \cite{fienup2003aberration, morrison2007sar}, and it is further demonstrated in \cite{morrison2007sar} that when the image is of multiple columns, and under certain statistical assumptions, the perfectly focused image can be closely approximated according to the strong law of large numbers. Image entropy is another popular alternative \cite{kragh2006monotonic, zeng2013sar,kantor2017minimum}, though in this case it is minimized (rather than maximized) to enforce sharpness. 

The second category of methods adopts an inverse problem approach. Based on a forward observation model associating the corrupted phase history with the underlying SAR image, SAR autofocusing is formulated as an inverse problem, and variational models with a variety of regularizers have been designed to obtain its solution. For instance, Onhon et al. uses the $p$th power of the approximate $l_{p}$ norm as the regularization term and an alternating minimization framework to solve the problem \cite{onhon2011sparsity}. There are also various methods addressing the problem in a compressive sensing context, such as the majorization-minimization based method \cite{kelly2012auto, kelly2014sparsity}, iteratively re-weighted augmented Lagrangian based method \cite{gungor2015augmented,  gungor2017autofocused} and conjugate gradient based method with a cost function involving hybrid regularization terms (approximate $l_{1}$ norm and approximate total variation regularization) \cite{ugur2012sar}. 

Besides these, there are also SAR autofocusing approaches built by directly strengthening traditional SAR imaging methods. For example, an autofocusing method which maximizes a sharpness metric for each pulse in the imaging process of back-projection is proposed in \cite{ash2011autofocus} and further extended to the case of moving ship targets \cite{sommer2019backprojection}. A polar format algorithm based autofocusing approach \cite{kantor2019polar} which combines \cite{onhon2011sparsity} with classical autofocusing method like PGA has been proposed recently as well. 

Moreover, with the development of deep learning, deep neural networks have also recently been considered in SAR autofocusing. A recurrent auto-encoder based SAR imaging network mimicking the behavior of iterative shrinkage thresholding algorithm (ISTA) is proposed \cite{mason2017deep}, in which the removal of phase errors is achieved by learning the forward model (observation matrix). Another auto-encoder and decoder based neural network is built in \cite{pu2021deep}, and the motion compensation is achieved by adding an updating step of the observation matrix in their alternating minimization framework.  

In this paper, we formulate the SAR autofocusing problem as an inverse problem, and adopt an alternating minimization framework to jointly estimate the desired SAR image and the unknown phase errors. This inverse problem poses an interesting challenge in comparison to many other inverse problems in imaging in that it deals with complex-valued functions.
Under this framework we will present two methods for SAR image formation and autofocusing. In both variants phase error estimation is achieved as in \cite{onhon2011sparsity}. However, for the sub-problem of estimating the SAR image, a Cauchy regularization on the magnitude of the desired image (thus we call it ``magnitude Cauchy") is used and we offer two different solving methods to handle its complex nature. The first solution is a complex-domain adaptation of the forward-backward splitting algorithm (CFBA). From a computational point of view, it transforms the problem of dealing with complex proximal operators into the problems of handling real proximal operators whose solutions are the magnitudes of the components of the originally desired complex solution. The second solution is based on Wirtinger calculus \cite{brandwood1983complex, van1994complex, kreutz2009complex, bouboulis2010wirtinger} which is designed to deal with differentiable (in the sense of \cite{brandwood1983complex}) real valued functions with complex variables. This method has already been introduced in our previous work \cite{zhang2021sar} and is referred to as Wirtinger alternating minimization autofocusing (WAMA) method, but we further give a thorough discussion of its convergence and show that it can be extended to the cases of several other regularizers.

The rest of this paper is organized as follows. In Section 2, a brief introduction is given to the data acquisition model for SAR and the associated problem formulation. In Section 3, the proposed forward-backward splitting based SAR autofocusing method is described in detail, then its convergence is analyzed.  In Section 4, the WAMA method is reviewed, and some more discussion are added, including its extension to the cases of other regularizers and its convergence. In Section 5, experimental results on both simulated scenes and real SAR images are shown to demonstrate the effectiveness of the proposed method. Finally, conclusions are presented in section 6.

\section{SAR Data acquisition model}
\label{sec:format}

In this paper, a SAR platform operating in spotlight mode is considered, whose transmitted signal at each azimuth position can be expressed as:
\begin{equation} \label{eq1}
	s(t)=Re\{e^{j(\omega_{0}t+\alpha t^{2})}\},
\end{equation}
where $\omega_{0}$ is the carrier frequency, 2$\alpha$ is the chirp rate, and $t$ is fast time. 

The obtained data $r_{m}(t)$ at the $m$th aperture position and the latent SAR image $F(x,y)$ can be associated by:
\begin{equation} \label{eq2}
	r_{m}(t)=\iint F(x,y)e^{-jU(xcos\theta+ysin\theta)}dxdy.
\end{equation}
The region over which the integral is computed is $x^{2}+y^{2}\le L^{2}$, with $L$ being the radius of the circular patch on the ground to be imaged. $\theta$ is the look angle, and $U$ is defined by
\begin{equation} \label{eq3}
	U=\frac{2}{c}(\omega_{0}+2\alpha(t-\tau_{0})),
\end{equation}
with $\tau_{0}$ being the demodulation time. The discretized version of this model is
\begin{equation} \label{eq4}
	\bm{r}_{m}=\bm{C}_{m}\bm{f},
\end{equation}
where $\bm{r}_{m}$ and $\bm{C}_{m}$ are the vector form of the phase history and the observation matrix for the $m$th aperture position respectively. $\bm{f}$ is the vector form of the underlying SAR image. Stacking (4) with respect to all the aperture positions, and considering phase errors as well as possible noise, the model becomes
\begin{equation} \label{eq5}
	\bm{g}=\bm{C}(\bm{\phi})\bm{f}+\bm{n},
\end{equation}
with $\bm{g}$ being the vector of corrupted phase history, $\bm{\phi}$ being the vector of phase errors, and $\bm{n}$ being the vector of Gaussian white noise. $\bm{C}(\bm{\phi})$ is the corrupted observation matrix. In this paper, only the case of 1D phase errors varying in the azimuth direction is dealt with, which leads to:
\begin{equation} \label{eq6}
	\bm{C}_{m}(\bm{\phi})=e^{j\bm{\phi}_{m}}\bm{C}_{m},
\end{equation}
where $\bm{C}_{m}(\bm{\phi})$ and $\bm{\phi}_{m}$ are the corrupted observation sub-matrix for the $m$th aperture position and the phase error for the $m$th aperture position respectively. The case of 2D separable phase errors as well as the case of 2D non-separable phase errors can be formulated similarly, see \cite{onhon2011sparsity}.

\section{The proposed CFBA method}

\subsection{The optimization model}
\label{sec:pagestyle}
We formulate SAR autofocusing as an inverse problem and minimize the following cost function:
\begin{equation} \label{oc}
	J(\bm{f},\bm{\phi})=\|\bm{g}-\bm{C}(\bm{\phi})\bm{f}\|_{2}^{2}-\lambda \sum_{i=1}^{N}\ln{\frac{\gamma}{\gamma^{2}+|\bm{f}_{i}|^{2}}},
\end{equation}
where $\lambda$ is the regularization parameter and $\gamma$ is the scale parameter for Cauchy distribution. 

The penalty term in (\ref{oc}) is a Cauchy regularization merely imposed on the magnitude of the latent SAR image. Consequently, we refer to it as ``magnitude Cauchy regularization". Like the $l_{p}$ norm, it too is a regularization term enforcing statistical sparsity \cite{McCullagh2018}, whose effectiveness has already been validated in SAR autofocusing \cite{zhang2021sar}, as well as SAR imaging and other inverse problems \cite{karakucs2020solving, karakucs2020convergence}.

In (\ref{oc}), the desired SAR image $\bm{f}$ and the phase errors $\bm{\phi}$ are both unknown. By jointly estimating them, SAR image reconstruction and the removal of phase errors are accomplished simultaneously. To do this, we adopt an alternating minimization autofocusing framework similar to \cite{onhon2011sparsity}. Specifically, $\bm{f}$ and $\bm{\phi}$ are updated alternatively by fixing one of them while optimizing the other. This iterative process will terminate when the relative error between $\bm{f}^{(n)}$ and $\bm{f}^{(n+1)}$ is smaller than $10^{-3}$.

\subsection{Complex forward-backward splitting based method}

\subsubsection{Image reconstruction step}

Under the framework of alternating minimization, each $\bm{f}$-sub-problem (\ref{fsub}) to be solved is 
\begin{equation} \label{fsub}
	\bm{f}^{(n+1)}=\mathrm{arg}\min_{\bm{f}\in \mathbb{C}^{N}}\|\bm{g}-\bm{C}(\bm{\phi}^{(n)})\bm{f}\|_{2}^{2}-\lambda \sum_{i=1}^{N}\ln{\frac{\gamma}{\gamma^{2}+|\bm{f}_{i}|^{2}}}.
\end{equation}

Unlike many other inverse problem formulations in computational imaging, (\ref{fsub}) is an optimization problem involving a complex unknown vector, and it thus needs to be handled with appropriate mathematical tools. To this end, we design an iterative solution, which is a complex version of the well-known forward-backward splitting algorithm, and thus we call it complex forward-backward autofocusing (CFBA). It holds the benefit that the computation of the involved complex proximity operators be converted to the computation of real proximity operators related to the magnitudes of original complex vector's components. Therefore, the techniques for real optimization regarding proximal operators can be leveraged. As a result, it is also possible to solve (\ref{oc}) with these techniques when the magnitude Cauchy regularization is replaced by certain non-smooth regularizers. Note that as will be discussed in Section 4, this is not always the case with the WAMA algorithm - the ability to be easily generalised to a number of non-smooth regularizers is a distinct advantage of the CFBA method.

Similar to the forward-backward splitting algorithm for real case, we recast (\ref{fsub}) as 
\begin{equation} \label{gph}
\bm{f}^{(n+1)}=\mathrm{arg}\min_{\bm{f}\in \mathbb{C}^{N}} H(\bm{f})+G(\bm{f}),
\end{equation}
where $H(\bm{f})=\|\bm{g}-\bm{C}(\bm{\phi}^{(n)})\bm{f}\|_{2}^{2}$ and $G(\bm{f})=\sum_{i=1}^{N}r(\bm{f}_{i})=-\lambda \sum_{i=1}^{N}\ln{\frac{\gamma}{\gamma^{2}+|\bm{f}_{i}|^{2}}}$. 

To minimize (\ref{gph}), with a given initial $\bm{o}^{(0)}$, we iteratively implement the following step:
\begin{equation} \label{fbi}
\bm{o}^{(k+1)}=\text{prox}_{\mu G}(\bm{o}^{(k)}-2\mu \nabla_{\bm{f}} H(\bm{o}^{(k)})),
\end{equation}
with
\begin{equation} \label{pr}
\text{prox}_{\mu G}(\bm{x})=\mathrm{arg}\min_{\bm{y}\in \mathbb{C}^{N}} \frac{1}{2}\|\bm{x}-\bm{y}\|_{2}^{2}+\mu G(\bm{y}).
\end{equation}
$\bm{f}^{(n+1)}$ is given by the final output $\bm{o}^{(K)}$ of this inner iterative loop. 

In (\ref{fbi}), $H(\bm{f})$ is a real-valued function with a complex vector variable, which is much discussed in Wirtinger calculus. Therefore, instead of using ordinary gradient operator defined in real case $\nabla$, here we use the complex gradient operator $\nabla_{\bm{f}}$ defined by Wirtinger calculus. Its definition is first proposed in \cite{brandwood1983complex} and further extended in \cite{kreutz2009complex}:
\begin{equation} \label{gwcg}
	\nabla_{\bm{f}}h=\bm{\Omega}_{\bm{f}}^{-1}(\frac{\partial h}{\partial \bm{f}})^{H}.
\end{equation}
where $\bm{\Omega}_{\bm{f}}^{-1}$ is a metric tensor. Using Brandwood’s setting, i. e., letting it be equal to the identity matrix, we have
\begin{equation} \label{bwcg}
	\nabla_{\bm{f}}h=(\frac{\partial h}{\partial \bm{f}})^{H}.
\end{equation}
As in (\ref{gwcg}) and (\ref{bwcg}), in the rest of this paper, we will use $\nabla_{\bm{f}}$ to denote the complex gradient operator defined by Wirtinger calculus, and $\nabla$ to denote the ordinary gradient operator defined in real case. 

Since the cost function is real-valued, according to \cite{kreutz2009complex}, we have:
\begin{equation} \label{wcrf}
	(\frac{\partial h}{\partial \bm{f}})^{H}=\overline{(\frac{\partial h}{\partial \bm{f}})}^{T}=(\frac{\partial h}{\partial \overline{\bm{f}}})^{T}=(\frac{\partial h}{\partial \overline{{\bm{f}}_{1}}},...,\frac{\partial h}{\partial \overline{{\bm{f}}_{N}}})^{T}.
\end{equation}

The last term in the right side of (\ref{wcrf}) can be computed using the chain rule and the definition of the conjugate $\mathbb{R}$-derivative, i. e., $\frac{\partial h}{\partial \overline{{\bm{f}}_{i}}}=\frac{1}{2}(\frac{\partial h}{\partial {\bm{x}}_{i}}+i\frac{\partial h}{\partial {\bm{y}}_{i}})$, with ${\bm{x}}_{i}$ and ${\bm{y}}_{i}$ being the real and imaginary part of ${\bm{f}}_{i}$ respectively \cite{kreutz2009complex}.

Also note that in (\ref{fbi}) the stepsize is written as $2\mu$ rather than the commonly used $\mu$ in the real-case forward-backward splitting algorithm. This choice is implied by the relationship between the complex gradient $\nabla_{\bm{f}}$ and real gradient $\nabla$, see e.g. \cite{brandwood1983complex}. Besides, $\mu$ should satisfy $\mu \le \frac{1}{L}$, as will be further discussed in Section 3.2.

As a result of Wirtinger calculus, it can be shown that
\begin{equation} \label{goH}
\nabla_{\bm{f}} H(\bm{f})=\bm{C}(\bm{\phi}^{(n)})^{H}(\bm{C}(\bm{\phi}^{(n)})\bm{f}-\bm{g}).
\end{equation}

As for the computation of (\ref{pr}), we first expand it as
\begin{equation} \label{pre}
\text{prox}_{\mu G}(\bm{x})=\mathrm{arg}\min_{\bm{y}\in \mathbb{C}^{N}} \frac{1}{2}\sum_{i=1}^{N}|\bm{x}_{i}-\bm{y}_{i}|^{2}-\mu\lambda \sum_{i=1}^{N}\ln{\frac{\gamma}{\gamma^{2}+|\bm{y}_{i}|^{2}}}.
\end{equation}

Then we solve (\ref{pre}) by independently solving
\begin{equation} \label{prm}
\text{prox}_{\mu \lambda r}(\bm{x}_{i})=\mathrm{arg}\min_{\bm{y}_{i}\in \mathbb{C}} \frac{1}{2}|\bm{x}_{i}-\bm{y}_{i}|^{2}-\mu \lambda \ln \frac{\gamma}{\gamma^{2}+|\bm{y}_{i}|^{2}}
\end{equation}
for each $i\ (i=1,...,N)$.

For (\ref{prm}), observe that the logarithm term in the Moreau envelope merely depends on $|\bm{y}_{i}|$, thus the solution $\bm{y}_{i}^{*}$ must lie on the line passing through the origin and the input $\bm{x}_{i}$, as long as $\bm{x}_{i}$ is not 0. Therefore if we fix $|\bm{y}_{i}|$, then only the first term of (\ref{prm}) needs to be minimized. To do this, we just need to find the point on a circle with radius $|\bm{y}_{i}|$ in the complex plane which is closest to the fixed point $\bm{x}_{i}$. Obviously, this point would be the one which also lies on the line passing through the origin and $\bm{x}_{i}$. Therefore, the desired point will have the same argument as $\bm{x}_{i}$. Since the choice of $|\bm{y}_{i}|$ is arbitrary, the argument of $\bm{y}_{i}^{*}$ must be the same as that of $\bm{x}_{i}$. 

However, if $\bm{x}_{i}=0$, every $\bm{y}_{i}$ on a certain circle of the complex plane will be the solution of (\ref{prm}). In this case, we set the argument of $\bm{y}_{i}$ as 0, as in \cite{soulez2016proximity}.

Therefore, the solution of (\ref{prm}) can now be split into two steps. The first is to solve the corresponding real optimization problem which gives $|\bm{y}_{i}^{*}|$:
\begin{equation} \label{rpom}
|\bm{y}_{i}^{*}|=\mathrm{arg}\min_{y \in \mathbb{R}} \frac{1}{2}(|\bm{x}_{i}|-y)^{2}-\mu \lambda \ln \frac{\gamma}{\gamma^{2}+y^{2}}.
\end{equation}
The second step is to let 
\begin{equation} \label{rpop}
\bm{y}_{i}^{*}=\left\{
\begin{array}{ll}
|\bm{y}_{i}^{*}|e^{j\phi_{\bm{x}_{i}}} & {\bm{x}_{i}\neq 0}\\
|\bm{y}_{i}^{*}| & {\bm{x}_{i}=0}
\end{array} \right.
\end{equation}
where $e^{j\phi_{\bm{x}_{i}}}=\bm{x}_{i}/|\bm{x}_{i}|$. Similar techniques can also be seen in phase retrieval \cite{soulez2016proximity} and SAR imaging \cite{guven2016augmented}. 

Note that (\ref{rpom}) is a non-convex optimization problem. If we compute the gradient of the Moreau envelope in the right-hand side and set it to 0, we will get a cubic equation. It may have three real roots, which stands for three stationary points. Here, however, we can add some constraints to simplify the problem. By restricting $\gamma \ge \frac{\sqrt{\mu\lambda}}{2}$, this Moreau envelope becomes convex and thus implies the existence of only one, real stationary point \cite{karakucs2020convergence}. In this case, the corresponding cubic equation must have a single real root and a pair of complex roots, and since our desired solution is a magnitude, then the solution we seek must be the real root. This real root is given by \cite{karakucs2020solving}:
\begin{equation} \label{cef}
|\bm{y}_{i}^{*}|=\frac{|\bm{x}_{i}|}{3}+s+t,
\end{equation}
where
\begin{equation} \label{ces}
s=\sqrt[3]{\frac{q}{2}+\sqrt{\frac{p^{3}}{27}+\frac{q^{2}}{4}}},
\end{equation}
\begin{equation} \label{cet}
t=\sqrt[3]{\frac{q}{2}-\sqrt{\frac{p^{3}}{27}+\frac{q^{2}}{4}}},
\end{equation}
\begin{equation} \label{cep}
p=\gamma^{2}+2\mu \lambda-\frac{|\bm{x}_{i}|^{2}}{3},
\end{equation}
\begin{equation} \label{ceq}
q=\gamma^{2}|\bm{x}_{i}|+\frac{2|\bm{x}_{i}|^{3}}{27}-\frac{\gamma^{2}+2\mu \lambda}{3}|\bm{x}_{i}|.
\end{equation}

\subsubsection{Optimization of the phase error}

After obtaining each $\bm{f}^{(n+1)}$, we use it to compute $\bm{\phi}^{(n+1)}$. For 1D phase error varying along the azimuth direction, the vector of phase errors can be updated by solving the following sequence of sub-problems concerning its components:

\begin{equation} \label{phisub}
	\bm{\phi}_{m}^{(n+1)}=\mathrm{arg}\min_{\bm{\phi}_{m} }\|\bm{g}_{m}-e^{j\bm{\phi}_{m}}\bm{C}_{m}\bm{f}^{(n+1)}\|_{2}^{2}, m=1,...,N,
\end{equation}
with $\bm{g}_{m}$ and $\bm{C}_{m}$ being the parts of $\bm{g}$ and $\bm{C}$ corresponding to the $m$th aperture position. According to \cite{onhon2011sparsity}, the solution is 
\begin{equation} \label{pss}
	\bm{\phi}_{m}^{(n+1)}=\arctan(\frac{\text{Re}\{[\bm{f}^{(n+1)}]^{H}\bm{C}_{m}\bm{g}_{m}\}}{\text{Im}\{[\bm{f}^{(n+1)}]^{H}\bm{C}_{m}\bm{g}_{m}\}}).
\end{equation}
The corrupted observation matrix can then be estimated by:
\begin{equation} \label{comu}
	\bm{C}_{m}(\bm{\phi}_{m}^{(n+1)})=e^{j\bm{\phi}_{m}^{(n+1)}}\bm{C}_{m}.
\end{equation}

The cases of 2D phase errors varying in both range direction and cross-range direction can also be solved by similar methods, see \cite{onhon2011sparsity} for more details. The whole process of the proposed CFBA method is summarized in Algorithm \ref{alg:example1} as follows:
\begin{algorithm}
\begin{algorithmic}
   \STATE{Initialize $n=0$, $\bm{f}^{(0)}=\bm{C}^{H}\bm{g}$, $\bm{\phi}^{(0)}=0$, $\bm{C}(\bm{\phi}^{0})=\bm{C}$, and set the values of $\gamma$, $\lambda$, and $\mu$ according to $\mu\in (0,\frac{1}{L})$ and $\gamma \ge \frac{\sqrt{\mu\lambda}}{2}$}
    \WHILE{$n<300$ or $\|\bm{f}^{(n+1)}-\bm{f}^{(n)}\|/\|\bm{f}^{(n)}\|>0.001$}
    \STATE{1. Compute $\bm{f}^{(n+1)}$ by complex forward-backward splitting method, i. e.:}
          \WHILE{$k<500$ or $\|\bm{o}^{(k+1)}-\bm{o}^{(k)}\|/\|\bm{o}^{(k)}\|>0.001$}
           \STATE{Find $\bm{o}^{(k+1)}=\text{prox}_{\lambda\mu R}(\bm{o}^{(k)}-\mu \bm{C}(\bm{\phi}^{(n)})^{H}(\bm{C}(\bm{\phi}^{(n)})\bm{o}^{(k)}-\bm{g}))$ by (\ref{rpom})-(\ref{ceq})}
           \STATE{$k=k+1$}
          \ENDWHILE
        \STATE{2. Compute $\phi_{m}^{(n+1)}$ by (\ref{pss})}
        \STATE{3. Compute $\bm{C}(\bm{\phi}_{m}^{(n+1)})$ by (\ref{comu})}
        \STATE{4. $n=n+1$}
    \ENDWHILE 
\end{algorithmic}
\caption{CFBA}
\label{alg:example1}
\end{algorithm}

\subsection{Convergence analysis}

For the proposed CFBA method, the issue of convergence is twofold. That is to say, the discussion needs to cover the convergence of the inner complex forward-backward splitting algorithm as well as the convergence of the outer alternating minimization algorithm.

\subsubsection{Convergence of the inner complex forward-backward splitting algorithm}

In the $n$th image reconstruction step, we find
\begin{equation} \label{8ag}
\bm{f}^{(n+1)}=\mathrm{arg}\min_{\bm{f}\in \mathbb{C}^{N}}\|\bm{g}-\bm{C}(\bm{\phi}^{(n)})\bm{f}\|_{2}^{2}-\lambda \sum_{i=1}^{N}\ln{\frac{\gamma}{\gamma^{2}+|\bm{f}_{i}|^{2}}},
\end{equation}
and in each step of the complex FB splitting algorithm which solves (\ref{8ag}) iteratively, we compute
\begin{equation} \label{fbiag}
\bm{o}^{(k+1)}=\mathrm{arg}\min_{\bm{o}\in \mathbb{C}^{N}} \frac{1}{2}\|\bm{o}-\bm{z}^{(k)}\|_{2}^{2}-\mu\lambda \sum_{i=1}^{N}\ln{\frac{\gamma}{\gamma^{2}+|\bm{o}_{i}|^{2}}},
\end{equation}
where
\begin{equation} \label{fboz}
\bm{z}^{(k)}=\bm{o}^{(k)}-2\mu \bm{C}(\bm{\phi}^{(n)})^{H}(\bm{C}(\bm{\phi}^{(n)})\bm{o}^{(k)}-\bm{g}).
\end{equation}

Since $\bm{\phi}^{(n)}$ is fixed for the $n$th $f$-sub-problem, for simplicity of notation, we will denote $\bm{C}(\bm{\phi}^{(n)})$ by $\bm{C}^{(n)}$ in this subsection. And using the notations of (\ref{gph}), we denote the cost function in (\ref{8ag}) as $J_{n}(\bm{f})=H(\bm{f})+G(\bm{f})$, with $H(\bm{f})=\|\bm{g}-\bm{C}^{(n)}\bm{f}\|_{2}^{2}$ and $G(\bm{f})=-\lambda \sum_{i=1}^{N}\ln{\frac{\gamma}{\gamma^{2}+|\bm{f}_{i}|^{2}}}$.

First we will prove that $J_{n}(\bm{f})$ is a real analytic function, and thus it satisfies Kurdyka-Lojasiewicz (KL) property \cite{attouch2010proximal}. For an arbitrary $N$-dimensional $\bm{f}=(\bm{f}_{1},...,\bm{f}_{N})^{T}=(\bm{x}_{1}+i\bm{y}_{1},...,\bm{x}_{N}+i\bm{y}_{N})^{T}\in \mathbb{C}^{N}$, we can obtain a $2N$-dimensional real vector $\widetilde{\bm{f}}=(\bm{x}_{1},...,\bm{x}_{N},\bm{y}_{1},...,\bm{y}_{N})^{T}\in \mathbb{R}^{2N}$. Conversely, for an arbitrary $2N$-dimensional $\widetilde{\bm{f}}=(\widetilde{\bm{f}}_{1},...,\widetilde{\bm{f}}_{N},\widetilde{\bm{f}}_{N+1},...,\widetilde{\bm{f}}_{2N})^{T}\in \mathbb{R}^{2N}$, we can obtain a $N$-dimensional complex vector $\bm{f}=(\bm{f}_{1},...,\bm{f}_{N})^{T}=(\widetilde{\bm{f}}_{1}+i\widetilde{\bm{f}}_{N+1},...,\widetilde{\bm{f}}_{N}+i\widetilde{\bm{f}}_{2N})^{T}\in \mathbb{C}^{N}$. For simplicity, we denote  $\widetilde{\bm{f}}_{R}=(\widetilde{\bm{f}}_{1},..., \widetilde{\bm{f}}_{N})^{T}$ and $\widetilde{\bm{f}}_{I}=(\widetilde{\bm{f}}_{N+1},..., \widetilde{\bm{f}}_{2N})^{T}$ for $\widetilde{\bm{f}}$. 

Now a plain but important fact is that $\|\bm{f}\|_{2}^{2}=\bm{f}^{H}\bm{f}=(\widetilde{\bm{f}})^{T}\widetilde{\bm{f}}=\|\widetilde{\bm{f}}\|_{2}^{2}$. Therefore, for $H(\bm{f})$, we have
\begin{equation} \label{Hf}
H(\bm{f})=\|\bm{g}-\bm{C}^{(n)}\bm{f}\|_{2}^{2}=\|\widetilde{\bm{g}}-\widetilde{\bm{C}^{(n)}\bm{f}}\|_{2}^{2}.
\end{equation}

Since
\begin{equation} \label{Cf1}
\bm{C}^{(n)}\bm{f}=(c_{11}\bm{f}_{1}+...+c_{1N}\bm{f}_{N},...,c_{N1}\bm{f}_{1}+...+c_{NN}\bm{f}_{N})^{T},
\end{equation}
we have
\begin{equation} \label{cftl}     
\widetilde{\bm{C}^{(n)}\bm{f}}=
\begin{pmatrix}   
(c_{11})_{R}\bm{x}_{1}-(c_{11})_{I}\bm{y}_{1}+...+(c_{1N})_{R}\bm{x}_{N}-(c_{1N})_{I}\bm{y}_{N}\\
\vdots\\
(c_{N1})_{R}\bm{x}_{1}-(c_{N1})_{I}\bm{y}_{1}+...+(c_{NN})_{R}\bm{x}_{N}-(c_{NN})_{I}\bm{y}_{N}\\
(c_{11})_{I}\bm{x}_{1}+(c_{11})_{R}\bm{y}_{1}+...+(c_{1N})_{I}\bm{x}_{N}+(c_{1N})_{R}\bm{y}_{N}\\
\vdots\\
(c_{N1})_{I}\bm{x}_{1}+(c_{N1})_{R}\bm{y}_{1}+...+(c_{NN})_{I}\bm{x}_{N}+(c_{NN})_{R}\bm{y}_{N}\\
\end{pmatrix},
\end{equation}
where $(c_{ij}),i=1,...,N, j=1,...,N$ are elements of $\bm{C}^{(n)}$, with $(c_{ij})_{R},i=1,...,N, j=1,...,N$ and $(c_{ij})_{I},i=1,...,N, j=1,...,N$ being the real and imaginary part of $(c_{ij})$ respectively.

And (\ref{cftl}) can be rewritten as
\begin{equation} \label{eqdcftl}
\widetilde{\bm{C}^{(n)}\bm{f}}=\widetilde{\bm{C}^{(n)}}\widetilde{\bm{f}},
\end{equation}
where
\begin{equation} \label{Cw}     
\widetilde{\bm{C}^{(n)}}=
\begin{pmatrix}   
(c_{11})_{R} & \cdots & (c_{1N})_{R} & (-c_{11})_{I} & \cdots & (-c_{1N})_{I} \\
\vdots & \ddots & \vdots & \vdots & \ddots & \vdots\\
(c_{N1})_{R} & \cdots & (c_{NN})_{R} & (-c_{N1})_{I} & \cdots & (-c_{NN})_{I} \\
(c_{11})_{I} & \cdots & (c_{1N})_{I} & (c_{11})_{R} & \cdots & (c_{1N})_{R} \\
\vdots & \ddots & \vdots & \vdots & \ddots & \vdots\\
(c_{N1})_{I} & \cdots & (c_{NN})_{I} & (c_{N1})_{R} & \cdots & (c_{NN})_{R} \\
\end{pmatrix}.
\end{equation}
As a result, $H(\bm{f})=\|\widetilde{\bm{g}}-\widetilde{\bm{C}^{(n)}}\widetilde{\bm{f}}\|_{2}^{2}=\widetilde{H}(\widetilde{\bm{f}})$. 

Now, $\widetilde{H}(\widetilde{\bm{f}})$ can be expressed as $\widetilde{H}(\widetilde{\bm{f}})=l(H_{(1)}(\widetilde{\bm{f}}),...H_{(2N)}(\widetilde{\bm{f}}))$, i. e., the composition of real analytic functions $l(\bm{x})=\bm{x}^{T}\bm{x}$ and $H_{(i)}(\widetilde{\bm{f}})=(\widetilde{\bm{g}})_{i}-(\widetilde{\bm{C}^{(n)}}\widetilde{\bm{f}})_{i}, i=1,..., 2N$. According to \cite{krantz2002primer}, $\widetilde{H}(\widetilde{\bm{f}})$ is a real analytic function of $\widetilde{\bm{f}}$.

On the other hand, for $G(\bm{f})$, we have
\begin{equation} \label{Gf}
\begin{split}
G(\bm{f})=-\sum_{i=1}^{N}\ln{\frac{\gamma}{\gamma^{2}+\|\widetilde{\bm{f}_{i}}\|^{2}}}=-\sum_{i=1}^{N}\ln{\frac{\gamma}{\gamma^{2}+\|\bm{s}_{(i)}\widetilde{\bm{f}}\|^{2}}}=\widetilde{G}(\widetilde{\bm{f}})
\end{split}
\end{equation}
where $\bm{s}_{(i)}$ is a $2\times 2N$ matrix whose first row and second row are the $i$th row and the $(i+N)$th row of a $2N\times 2N$ identity matrix respectively. Therefore, each summed term of $\widetilde{G}(\widetilde{\bm{f}})$, i.e, $-\ln{\frac{\gamma}{\gamma^{2}+\|\bm{s}_{(i)}\widetilde{\bm{f}}\|^{2}}}$ can be written as $-\ln{\frac{\gamma}{\gamma^{2}+l(G_{(1)}(\widetilde{\bm{\bm{f}}}), G_{(2)}(\widetilde{\bm{f}}))}}$, where $l(\bm{x})=\bm{x}^{T}\bm{x}$ and $G_{(j)}(\widetilde{\bm{f}})=(\bm{s}_{(i)}\widetilde{\bm{f}})_{j}, j=1, 2$. That is to say, it is a composition of real analytic functions, thus is itself real analytic. As a result, $\widetilde{G}(\widetilde{\bm{f}})$ is a real analytic function of $\widetilde{\bm{f}}$.

Based on the above two conclusions, $J_{n}(\bm{f})=H(\bm{f})+G(\bm{f})=\widetilde{H}(\widetilde{\bm{f}})+\widetilde{G}(\widetilde{\bm{f}})=\widetilde{J_{n}}(\widetilde{\bm{f}})$ is a real analytic function of $\widetilde{\bm{f}}$, and thus satisfies the KL property. This means that for every $\widetilde{\bm{f}}^{'}\in \mathbb{R}^{2N}$ and every bounded neighborhood $U$ of $\widetilde{\bm{f}}^{'}$, there exists $\kappa \in (0,+\infty)$, $\eta \in (0,+\infty)$ and $\theta \in [0,1)$ such that
\begin{equation} \label{KL}
\|\nabla \widetilde{J_{n}}(\widetilde{\bm{f}})\|\ge \kappa|\widetilde{J_{n}}(\widetilde{\bm{f}})-\widetilde{J_{n}}(\widetilde{\bm{f}}^{'})|^{\theta}
\end{equation}
for every $\widetilde{\bm{f}}\in U\cap\{\widetilde{\bm{f}}||\widetilde{J_{n}}(\widetilde{\bm{f}})-\widetilde{J_{n}}(\widetilde{\bm{f}}^{'})|\le \eta\}$.

This proof of the KL property also implies the proof of the convergence of CFBA algorithm from a perspective of real vector variables, because $J_{n}(\bm{f})$ (a function of the complex $\bm{f}$) can now be viewed as $\widetilde{J_{n}}(\widetilde{\bm{f}})$ (a function of real vector variable $\widetilde{\bm{f}}$). If the real forward-backward splitting algorithm minimizing $\widetilde{J_{n}}(\widetilde{\bm{f}})$ can be proven to be equivalent to the proposed complex forward-backward splitting algorithm minimizing $J_{n}(\bm{f})$, then the convergence analysis of the latter is equivalent to the convergence analysis of the former.

Let us first formulate the real forward-backward splitting algorithm minimizing $\widetilde{J}_{n}(\widetilde{\bm{f}})$ with respect to $\widetilde{\bm{f}}$. First denote $\bm{u}^{(0)}=\widetilde{\bm{o}^{(0)}}$. Then in each step, this algorithm finds
\begin{equation} \label{rfbit}
\begin{split}
\bm{u}^{(k+1)}=\text{prox}_{\mu \widetilde{G}}(\bm{w}^{(k)})=\mathrm{arg}\min_{\bm{u}\in \mathbb{R}^{2N}} \frac{1}{2}\|\bm{u}-\bm{w}^{(k)}\|_{2}^{2}-\mu \lambda \sum_{i=1}^{N}\ln{\frac{\gamma}{\gamma^{2}+\|\bm{s}_{(i)}\bm{u}\|^{2}}},
\end{split}
\end{equation}
where
\begin{equation} \label{rfbw}
\bm{w}^{(k)}=\bm{u}^{(k)}-2\mu\widetilde{\bm{C}^{(n)}}^{T}(\widetilde{\bm{C}^{(n)}}\bm{u}^{(k)}-\widetilde{\bm{g}}),
\end{equation}

Observe that
\begin{equation} \label{rC1}
(\bm{C}^{(n)})^{H}\bm{f}=(\overline{c_{11}}\bm{f}_{1}+...+\overline{c_{N1}}\bm{f}_{N},...,\overline{c_{1N}}\bm{f}_{1}+...+\overline{c_{NN}}\bm{f}_{N})^{T},
\end{equation}
and
\begin{equation} \label{rC2}     
\widetilde{(\bm{C}^{(n)})^{H}\bm{f}}=
\begin{pmatrix}   
(c_{11})_{R}\bm{x}_{1}+(c_{11})_{I}\bm{y}_{1}+...+(c_{N1})_{R}\bm{x}_{N}+(c_{N1})_{I}\bm{y}_{N}\\
\vdots\\
(c_{1N})_{R}\bm{x}_{1}+(c_{1N})_{I}\bm{y}_{1}+...+(c_{NN})_{R}\bm{x}_{N}+(c_{NN})_{I}\bm{y}_{N}\\
(-c_{11})_{I}\bm{x}_{1}+(c_{11})_{R}\bm{y}_{1}+...+(-c_{N1})_{I}\bm{x}_{N}+(c_{N1})_{R}\bm{y}_{N}\\
\vdots\\
(-c_{1N})_{I}\bm{x}_{1}+(c_{1N})_{R}\bm{y}_{1}+...+(-c_{NN})_{I}\bm{x}_{N}+(c_{NN})_{R}\bm{y}_{N}\\
\end{pmatrix}.
\end{equation}

This can be exactly decomposed as
\begin{equation} \label{rC3}
\widetilde{(\bm{C}^{(n)})^{H}\bm{f}}=\widetilde{\bm{C}^{(n)}}^{T}\widetilde{\bm{f}},
\end{equation}
where $\widetilde{\bm{C}^{(n)}}$ is as in (\ref{Cw}).

Therefore, we have
\begin{equation} \label{rw0}
\bm{w}^{(0)}=\widetilde{\bm{o}^{(0)}}-2\mu\widetilde{\bm{C}^{(n)}}^{T}(\widetilde{\bm{C}^{(n)}}\widetilde{\bm{o}^{(0)}}-\widetilde{\bm{g}})=\widetilde{\bm{o}^{(0)}}-2\mu\widetilde{\nabla h(\bm{o}^{(0)})}=\widetilde{\bm{z}^{(0)}},
\end{equation}
with $\bm{z}^{(0)}$ defined by (\ref{fboz}) for CFBA. 

If now we rewrite the $\bm{u}$ in (\ref{rfbit}) by $\bm{u}=\widetilde{\bm{o}}$, then the Moreau envelope therein becomes a function of $\widetilde{\bm{o}}$. And if this function is rewritten as a function of $\bm{o}$, the result will exactly take the form of the Moreau envelope in (\ref{fbiag}). Moreover, we have $\nabla \widetilde{J_{n}}(\widetilde{\bm{f}})=((\frac{\partial \widetilde{J_{n}}}{\partial \widetilde{\bm{f}}_{R}})^{T}, (\frac{\partial \widetilde{J_{n}}}{\partial \widetilde{\bm{f}}_{I}})^{T})^{T}$ by the definition of real gradient and $\nabla_{\bm{f}} J_{n}(\bm{f})=\frac{1}{2}(\frac{\partial \widetilde{J_{n}}}{\partial \widetilde{\bm{f}}_{R}}+i\frac{\partial \widetilde{J_{n}}}{\partial \widetilde{\bm{f}}_{I}})$ by the definition in Wirtinger calculus. Therefore, if $\bm{f}$ is a stationary point of $J_{n}(\bm{f})$, then the corresponding $\widetilde{\bm{f}}$ is a stationary point of $\widetilde{J_{n}}(\widetilde{\bm{f}})$. Besides, since we have forced convexity of the Moreau envelopes in (\ref{rfbit}) and (\ref{fbiag}) by restricting the range of parameters, they will each have only one stationary point. Due to these three conclusions, $\bm{u}^{(1)}=\widetilde{\bm{o}^{(1)}}$ holds. 

By induction, following similar deduction, a sequence $\bm{o}^{(k)}$ complying with (29) in CFBA algorithm and meanwhile satisfying $\bm{u}^{(k)}=\widetilde{\bm{o}^{(k)}}$ for all $k$ can be obtained. Therefore, the convergence of CFBA algorithm can be analyzed equivalently by discussing this real FB splitting algorithm.

The analysis above also implies that the $n$th $f$-sub-problem can be solved by finding the corresponding real solution and transforming it back to the desired complex solution. However, the proposed CFBA algorithm is more compact in form, since it deals with $N$ dimensional vectors instead of $2N$ dimensional vectors, and it doesn't require the construction of $\widetilde{C}$ based on $C$.  

Now, $\widetilde{J}_{n}(\widetilde{\bm{f}})=\widetilde{G}(\widetilde{\bm{f}})+\widetilde{H}(\widetilde{\bm{f}})$ is proper, lower semicontinuous, bounded from below, and satisfies the KL property. $\widetilde{H}(\widetilde{\bm{f}})$ is finite valued, differentiable, and has a Lipschitz continuous gradient. Moreover $\widetilde{G}(\widetilde{\bm{f}})$ is continuous on its domain. That is to say, all the conditions in theorem 5.1 of \cite{attouch2013convergence} are satisfied. Therefore, according to that theorem, with the conditions aforementioned and the setting $\mu \le \frac{1}{L}$ (which guarantees the monotonic decreasing nature of $\widetilde{J}_{n}(\bm{u}^{(k)})$), we come to the conclusion that the iterates $\bm{u}^{(k)}$ will converge to some critical point of $\widetilde{J}_{n}(\widetilde{\bm{f}})$. Equivalently, the iterates $\bm{o}^{(k)}$ produced by the proposed CFBA method will converge to some critical point of $J_{n}(\bm{f})$.

Moreover, let us denote by $\widetilde{\bm{f}}^{*}$ the global minimizer of $\widetilde{J}_{n}(\widetilde{\bm{f}})$. According to Theorem 2.12 in \cite{attouch2013convergence}, for each $r>0$, there exist $u\in (0,r), \delta>0$ such that the inequalities $\|\bm{u}^{(0)}-\widetilde{\bm{f}}^{*}\|<u$ and $\min \widetilde{J}_{n}(\widetilde{\bm{f}})<\widetilde{J}_{n}(\bm{u}^{(0)})<\delta+\min \widetilde{J}_{n}(\widetilde{\bm{f}})$ imply that the sequence $\bm{u}^{(k)}$ generated for each $\widetilde{J}_{n}(\widetilde{\bm{f}})$ will converge to some ${\bm{u}}^{*}$ with $\bm{u}^{(k)}\in B(\widetilde{\bm{f}}^{*}, r)$ for arbitrary $k$ and $\widetilde{J}_{n}(\bm{u}^{*})=\min \widetilde{J}_{n}(\widetilde{\bm{f}})$. That is to say, convergence of the sequence $\bm{u}^{(k)}$ to a global minimizer of $\widetilde{J}_{n}(\widetilde{\bm{f}})$ can be obtained. Equivalently, the iterates $\bm{o}^{(k)}$ produced by the proposed CFBA method will converge to a global minimizer of $J_{n}(\bm{f})$.

In fact, the above proof which is elaborated from a real perspective can also be done alternatively by working on the complex iterates $\bm{o}^{(k)}$ themselves. However, this process is more complicated (see Appendix A).

\subsubsection{Convergence of the outer alternating minimization method}

For each $\bm{f}$-sub-problem, if the assumptions related to the initial value $\bm{u}^{(0)}$ stated in the last section are satisfied, the corresponding sequence $\bm{o}^{(k)}$ will converge to a global minimizer of $J_{n}(\bm{f})$. In this case, it is reasonable to assert that $J_{n}(\bm{f}^{(n+1)},\bm{\phi}^{(n)})\le J_{n}(\bm{f}^{(n)},\bm{\phi}^{(n)})$. And since each $\phi$-sub-problem has a closed form solution, we have $J_{n}(\bm{f}^{(n+1)},\bm{\phi}^{(n+1)})\le J_{n}(\bm{f}^{(n+1)},\bm{\phi}^{(n)})$. As a result, $J_{n}(\bm{f}^{(n+1)},\bm{\phi}^{(n+1)})\le J_{n}(\bm{f}^{(n)},\bm{\phi}^{(n)})$ holds for every $n$, i. e., $J(\bm{f}^{(n)},\bm{\phi}^{(n)})$ is a monotonically decreasing sequence. Since it is also bounded below, it will converge to a certain value, though not necessarily equal to $\inf J(\bm{f},\bm{\phi})$.

Besides, if stronger assumptions are satisfied, better results of convergence can be obtained. For instance, if the five-point property \cite{csiszar1984information,byrne2013alternating} holds, i. e., if
\begin{equation} \label{fpp}
J(\bm{f},\bm{\phi})+J(\bm{f},\bm{\phi}^{(n)})\ge J(\bm{f},\bm{\phi}^{(n+1)})+J(\bm{f}^{(n+1)},\bm{\phi}^{(n)})
\end{equation}
holds for every $\bm{f}, \bm{\phi}$, and $n$, then
\begin{equation} \label{fppc}
\lim_{n \to +\infty}J(\bm{f}^{(n)},\bm{\phi}^{(n)})=\inf J(\bm{f},\bm{\phi}).
\end{equation}

\section{Wirtinger alternating minimization autofocusing}

\subsection{The original method}

In this section, we review the Wirtinger alternating minimization autofocusing (WAMA) method originaly proposed in \cite{zhang2021sar}. After that, we briefly expand on how to extend this method to several other cost functions with the same fidelity term but with different regularizers. Finally, we will discuss the convergence of this method, a topic not covered in previous publications.

WAMA method also adopts the framework of alternating minimization, including two types of sub-problems to be solved. For each $\bm{f}$-sub-problem formulated in (\ref{fsub}), Wirtinger calculus is used to solve it. On the one hand, Wirtinger calculus is a powerful theory covering the analysis of real-valued functions of complex variables, and within which many real optimization problems can have their complex counterparts defined. On the other hand, it is also a rather elegant approach, due to its ability to address the problem in a concise way. Namely, there is no need to expand the complex variables as real vectors in the computational process.

Specifically, to solve (\ref{fsub}), we compute the complex gradient of the cost function therein directly using Wirtinger calculus. For the second term of the cost function, if we denote
\begin{equation} \label{ctorf}
	R(\bm{f})=-\sum_{i=1}^{N}\ln{\frac{\gamma}{\gamma^{2}+|\bm{f}_{i}|^{2}}},
\end{equation}
then we have
\begin{equation} \label{gorf}
	(\nabla_{\bm{f}} R(\bm{f}))_{i}=\frac{\bm{f}_{i}}{\gamma^{2}+|\bm{f}_{i}|^{2}},  i=1,...,N.
\end{equation}

As for the first term, using the results in \cite{brandwood1983complex}, it is obvious that:
\begin{equation} \label{gft}
\nabla_{\bm{f}} \|\bm{g}-\bm{C}(\bm{\phi}^{(n)})\bm{f}\|_{2}^{2}=\bm{C}(\bm{\phi}^{(n)})^{H}(\bm{C}(\bm{\phi}^{(n)})\bm{f}-\bm{g}).
\end{equation}

Therefore, the complex gradient of (\ref{fsub}) can be written as:
\begin{equation} \label{gfsub}
\nabla_{\bm{f}} J(\bm{f},\bm{\phi})=\bm{C}(\bm{\phi}^{(n)})^{H}(\bm{C}(\bm{\phi}^{(n)})\bm{f}-\bm{g})+\lambda \bm{W}(\bm{f})\bm{f},
\end{equation}
where 
\begin{equation} \label{mtw}
\bm{W}(\bm{f})=\text{diag}(\bm{s}),
\end{equation}
\begin{equation} \label{sinw}
\bm{s}_{i}=\frac{1}{\gamma^{2}+|\bm{f}_{i}|^{2}}, i=1,...,N.
\end{equation}

Now we set (\ref{gfsub}) to zero, according to the necessary and sufficient condition for a stationary point of a real-valued complex function \cite{brandwood1983complex, kreutz2009complex}, which leads to
\begin{equation} \label{fsubou}
	[\bm{C}(\bm{\phi}^{(n)})^{H}\bm{C}(\bm{\phi}^{(n)})+\lambda \bm{W}(\bm{f})]\bm{f}=\bm{C}(\bm{\phi}^{(n)})^{H}\bm{g}.
\end{equation}
It is worth pointing out that even though the exact solution of (\ref{fsubou}) can be obtained, that solution is not necessarily the global minimum of (\ref{fsub}) due to the non-convexity of the Cauchy penalty. 

Now we rewrite (\ref{fsubou}) as $\bm{A}\bm{f}=\bm{b}$, where $\bm{b}=\bm{C}(\bm{\phi}^{(n)})^{H}\bm{g}$ and $\bm{A}=\lambda \bm{W}(\bm{f})+\bm{C}(\bm{\phi}^{(n)})^{H}\bm{C}(\bm{\phi}^{(n)})$. 
Since $\bm{W}(\bm{f})$ depends on $\bm{f}$, so does $\bm{A}$. This makes (\ref{fsubou}) nonlinear in respect to $\bm{f}$, and it is difficult to find its closed form solution. However, if we sacrifice some accuracy and approximate the $\bm{f}$ in $\bm{W}(\bm{f})$ with the $\bm{f}$ computed during the last iteration of the alternating minimization framework, $\bm{A}$ is converted into a constant matrix, and thus $\bm{A}\bm{f}=\bm{b}$ becomes a linear system of equations which can be solved efficiently. 

That is to say, when computing an unknown $\bm{f}^{(n+1)}$, the actually solved equation is
\begin{equation} \label{fsubau}
	[\bm{C}(\bm{\phi}^{(n)})^{H}\bm{C}(\bm{\phi}^{(n)})+\lambda \bm{W}(\bm{f}^{(n)})]\bm{f}^{(n+1)}=\bm{C}(\bm{\phi}^{(n)})^{H}\bm{g}.
\end{equation}
This equation can be viewed as a fixed-point algorithm with a single iterative step, and its solution can be efficiently obtained by using the conjugate gradient (CG) algorithm \cite{barrett1994templates}. The experimental results in Section 5 imply that the obtained solution is sufficiently good.

As for the $\bm{\phi}$-sub-problems, the solutions are the same as that of CFBA introduced in 3.1, so they are not presented here. Now, the whole process of WAMA method can be summarized as Algorithm \ref{alg:example2} as follows:

\begin{algorithm}
\begin{algorithmic}
   \STATE{Initialize $n=0$, $\bm{f}^{(0)}=\bm{C}^{H}\bm{g}$, $\bm{\phi}^{(0)}=0$, $\bm{C}(\bm{\phi}^{0})=\bm{C}$, and set the values of $\gamma$ and $\lambda$}
    \WHILE{$n<300$ or $\|\bm{f}^{(n+1)}-\bm{f}^{(n)}\|/\|\bm{f}^{(n)}\|>0.001$}
        \STATE{1. Compute $\bm{f}^{(n+1)}$ by finding the solution of (\ref{fsubau}) via CG}
        \STATE{2. Compute $\bm{\phi}_{m}^{(n+1)}$ by (\ref{pss})}
        \STATE{3. Compute $\bm{C}(\bm{\phi}_{m}^{(n+1)})$ by (\ref{comu})}
        \STATE{4. $n=n+1$}
    \ENDWHILE 
\end{algorithmic}
\caption{WAMA}
\label{alg:example2}
\end{algorithm}

\subsection{Extension to several other regularizers}

As an extension, the same computational processes can also be followed to handle the cases where the magnitude Cauchy regularization in (\ref{oc}) is replaced by some other $\mathbb{R}$-differentiable regularizers \cite{kreutz2009complex}. For those cases, equation (\ref{fsubau}) will also be obtained, despite the fact that the involved $\bm{s}$ is different. We give several examples as follows:

(1) {\it $p$th power of approximate $l_{p}$ norm}

In this case,
\begin{equation} \label{eqalr}
R(\bm{f})=\sum_{i=1}^{N}(|\bm{f}_{i}|^{2}+\beta)^{\frac{p}{2}},
\end{equation}
and now
\begin{equation} \label{eqals}
\bm{s}_{i}=\frac{p}{2(|\bm{f}_{i}|^{2}+\beta)^{1-\frac{p}{2}}}, i=1,...,N.
\end{equation}
We point out that the updating formula in this case is the same as that in \cite{onhon2011sparsity}, but no reference to the literature of Wirtinger calculus is made there.

(2) {\it approximate total variation}

For approximate total variation, the situation is more complicated. However, the result can still be incorporated in the form of (\ref{fsubau}). Let $\bm{F}$ be the 2D $a\times b$ matrix form of the $N$-dimensional vector $\bm{f} (N=a\times b)$, then
\begin{equation} \label{eqatv}
R(\bm{f})=\sum_{i=1}^{N}\sum_{j=1}^{N}\sqrt{|(\nabla_{i}\bm{F})_{i,j}|^{2}+|(\nabla_{j}\bm{F})_{i,j}|^{2}+\beta},
\end{equation}
with
\begin{equation} \label{eqdf1}
(\nabla_{i}\bm{F})_{i,j}=\left\{
\begin{array}{ll}
\bm{F}_{i,j}-\bm{F}_{i-1,j} & {i>1}\\
0 & {i=1}
\end{array} \right.
\end{equation}
\begin{equation} \label{eqdf2}
(\nabla_{j}\bm{F})_{i,j}=\left\{
\begin{array}{ll}
\bm{F}_{i,j}-\bm{F}_{i,j-1} & {j>1}\\
0 & {j=1}
\end{array} \right.
\end{equation}
And now 
\begin{equation} \label{eqwtv}
\bm{W}(\bm{f})=\bm{W}^{'}\bm{D}^{'}+\bm{W}^{'}\bm{D}^{''}+\bm{W}^{''}\bm{D}^{'''}+\bm{W}^{'''}\bm{D}^{''''},
\end{equation}
\begin{equation} \label{eqwtv1}
\bm{W}^{'}=\text{diag}(\text{vec}(\bm{S}^{'})), \bm{W}^{''}=\text{diag}(\text{vec}(\bm{S}^{''})), \bm{W}^{'''}=\text{diag}(\text{vec}(\bm{S}^{'''})).
\end{equation}
where $\text{vec}$ is the operation which turns a matrix into a column vector by stacking its columns in order. And
\begin{equation} \label{eqstv1}
(\bm{S}^{'})_{i,j}=\frac{1}{2\sqrt{|(\nabla_{i}\bm{F})_{i,j}|^{2}+|(\nabla_{j}\bm{F})_{i,j}|^{2}+\beta}},
\end{equation}
\begin{equation} \label{eqstv2}
(\bm{S}^{''})_{i,j}=\frac{1}{2\sqrt{|(\nabla_{i}\bm{F})_{i,j+1}|^{2}+|(\nabla_{j}\bm{F})_{i,j+1}|^{2}+\beta}},
\end{equation}
\begin{equation} \label{eqstv3}
(\bm{S}^{'''})_{i,j}=\frac{1}{2\sqrt{|(\nabla_{i}\bm{F})_{i+1,j}|^{2}+|(\nabla_{j}\bm{F})_{i+1,j}|^{2}+\beta}}.
\end{equation}
As for $\bm{D}^{'}, \bm{D}^{''}, \bm{D}^{'''}$, and $\bm{D}^{''''}$, they are matrices contains only 0, 1, and -1, and constructed so that they realize the following relations:
\begin{equation} \label{eqdtv1}
\begin{split}
\bm{D}^{'}\bm{f}=\text{vec}((\nabla_{i}\bm{F})_{i,j}),\  \bm{D}^{''}\bm{f}=\text{vec}((\nabla_{j}\bm{F})_{i,j}),\\
\bm{D}^{'''}\bm{f}=-\text{vec}((\nabla_{j}\bm{F})_{i,j+1}),\ 
\bm{D}^{''''}\bm{f}=-\text{vec}((\nabla_{i}\bm{F})_{i+1,j}).
\end{split}
\end{equation}

(3) {\it Welsh potential}

In this case, a $l_{2}-l_{0}$ regularization \cite{florescu2014majorize} is imposed on the magnitude of $f$, and we have
\begin{equation} \label{eql0w}
R(\bm{f})=\sum_{i=1}^{N}(1-e^{-\frac{|\bm{f}_{i}|^{2}}{2\delta^{2}}}),
\end{equation}
and
\begin{equation} \label{eql0ws}
\bm{s}_{i}=\frac{e^{-\frac{|\bm{f}_{i}|^{2}}{2\delta^{2}}}}{2\delta^{2}}, i=1,...,N.
\end{equation}

(4) {\it Geman-McClure potential} 

In this case, another variant of $l_{2}-l_{0}$ regularization \cite{florescu2014majorize} is imposed on the magnitude of $f$, and we have
\begin{equation} \label{eql0gm}
R(\bm{f})=\sum_{i=1}^{N}\frac{|\bm{f}_{i}|^{2}}{2\delta^{2}+|\bm{f}_{i}|^{2}},
\end{equation}
and
\begin{equation} \label{eql0gms}
\bm{s}_{i}=\frac{2\delta^{2}}{(2\delta^{2}+|\bm{f}_{i}|^{2})^{2}}, i=1,...,N.
\end{equation}

\subsection{Convergence analysis}
The approximation (\ref{fsubau}) used for the solution of each image reconstruction step adds much difficulty to the discussion of the convergence of WAMA method. However, the WAMA method can be analyzed from another perspective, which renders its convergence analysis tractable. 

Similar to \cite{ccetin2006feature}, the key point is the  construction of a $K(\bm{b},\bm{f},\bm{\phi})$ such that $\inf_{\bm{b}}{K(\bm{b},\bm{f},\bm{\phi})}=J(\bm{f},\bm{\phi})$, with $J(\bm{f},\bm{\phi})$ given by (\ref{oc}). Following the theories introduced in \cite{geman1992constrained}, this $K(\bm{b},f,\bm{\phi})$ is constructed as
\begin{equation} \label{af}
K(\bm{b},\bm{f},\bm{\phi})=\|\bm{g}-\bm{C}(\bm{\phi})\bm{f}\|_{2}^{2}-\lambda\sum_{i=1}^{N}[(|\bm{f}_{i}|^{2}+\gamma^{2})\bm{b}_{i}-\ln{(\gamma\bm{b}_{i})}-1],
\end{equation}
where $\bm{b}$ is an auxiliary vector.

Now, for verification, we let $\frac{\partial{K}}{\partial{\bm{b}}}=0$ to find the $\bm{b}^{*}$ minimizing $K(\bm{b},f,\bm{\phi})$ for a fixed $\bm{f}$ and $\bm{\phi}$. Consequently,
\begin{equation} \label{bmc}
\bm{b}_{i}^{*}=\frac{1}{\gamma^{2}+|\bm{f}_{i}|^{2}}.
\end{equation}
Substituting (\ref{bmc}) into (\ref{af}), $K(\bm{b}^{*},\bm{f},\bm{\phi})=J(\bm{f},\bm{\phi})$ is obtained, and therefore the equality $\inf_{\bm{b}}{K(\bm{b},\bm{f},\bm{\phi})}=J(\bm{f},\bm{\phi})$ is verified.

Therefore, minimizing the original cost function (\ref{oc}) with respect to $\bm{f}$ and $\bm{\phi}$ is equivalent to minimizing (\ref{af}) with respect to $\bm{b}$, $\bm{f}$, and $\bm{\phi}$. If an alternating minimization scheme is imposed directly on $K(\bm{b},\bm{f},\bm{\phi})$, the procedure will consist of the repetition of the following three steps:

1. Find $\bm{b}^{(n+1)}$ by
\begin{equation} \label{bm}
\bm{b}^{(n+1)}=\mathrm{arg}\min_{\bm{b}}K(\bm{b},\bm{f}^{(n)},\bm{\phi}^{(n)}).
\end{equation}
This leads to:
\begin{equation} \label{bs}
\bm{b}_{i}^{(n+1)}=\frac{1}{\gamma^{2}+|\bm{f}_{i}^{(n)}|^{2}}.
\end{equation}

2. Find $\bm{f}^{(n+1)}$ by
\begin{equation} \label{fm}
\bm{f}^{(n+1)}=\mathrm{arg}\min_{\bm{f}}K(\bm{b}^{(n+1)},\bm{f},\bm{\phi}^{(n)}).
\end{equation}
This lead to:
\begin{equation} \label{fs}
[\bm{C}(\bm{\phi}^{(n)})^{H}\bm{C}(\bm{\phi}^{(n)})+\lambda W]\bm{f}^{(n+1)}=\bm{C}(\bm{\phi}^{(n)})^{H}\bm{g},
\end{equation}
where 
\begin{equation} \label{kfw}
W=\text{diag}(\bm{b}^{(n+1)}),
\end{equation}

3. Find $\bm{\phi}^{(n+1)}$ by
\begin{equation} \label{pm}
\bm{\phi}^{(n+1)}=\mathrm{arg}\min_{\bm{\phi}}K(\bm{b}^{(n+1)},\bm{f}^{(n+1)},\bm{\phi}).
\end{equation}
This leads to:
\begin{equation} \label{ps}
\bm{\phi}_{m}^{(n+1)}=\arctan(\frac{\text{Re}\{[\bm{f}^{(n+1)}]^{H}\bm{C}_{m}\bm{g}_{m}\}}{\text{Im}\{[\bm{f}^{(n+1)}]^{H}\bm{C}_{m}\bm{g}_{m}\}}).
\end{equation}

Notice that if we combine (\ref{bs}) and (\ref{fs}) as one step, then the formulas (\ref{bs}), (\ref{fs}) and (\ref{ps}) are exactly the same as (\ref{fsubau}) and (\ref{pss}). Therefore, the convergence of the WAMA method can be analyzed by discussing this equivalent alternating minimization process.

According to WAMA, the conjugate gradient (CG) method is used here to obtain the solution of (\ref{fs}). Denote by $\bm{f}^{*(n)}$ the exact solution of (\ref{fs}), and by $\bm{q}^{(j)}$ ($j=1, ..., J$) the iterates in the loop of CG such that $\bm{f}^{(n+1)}=\bm{q}^{(J)}$. According to \cite{joly1993complex}, if the matrix $\bm{A}=[\bm{C}(\bm{\phi}^{(n)})^{H}\bm{C}(\bm{\phi}^{(n)})+\lambda \bm{W}]$ is non-singular, we can get
\begin{equation} \label{ccg}
\|\bm{q}^{(j)}-\bm{f}^{*(n)}\|_{\bm{A}}^{2}\le \|\bm{q}^{(0)}-\bm{f}^{*(n)}\|_{\bm{A}}^{2}(\frac{\sqrt{cond(\bm{A})}-1}{\sqrt{cond(\bm{A})}+1})^{2j},
\end{equation}
with $\|\bm{x}\|_{\bm{A}}^{2}=\bm{x}^{H}\bm{A}\bm{x}$, and $cond(\bm{A})$ being the condition number of $\bm{A}$. That is to say, each set of iterates $\bm{q}^{(j)}$ generated by the conjugate gradient method will converge to its corresponding $\bm{f}^{*(n)}$ as $j$ goes to infinity.

Therefore, since (\ref{bs}), (\ref{fs}), (\ref{ps}) all give closed-form solutions or sufficient accuracy (suggested by the convergence analysis, see (\ref{ccg})), we assert that:
\begin{equation} \label{deK}
K(\bm{b}^{(n+1)},\bm{f}^{(n+1)},\bm{\phi}^{(n+1)})\le K(\bm{b}^{(n)},\bm{f}^{(n)},\bm{\phi}^{(n)}).
\end{equation}
That is to say, $K(\bm{b}^{(n)},\bm{f}^{(n)},\bm{\phi}^{(n)})$ is a monotonically decreasing sequence. And since it is bounded below, it will converge to a certain value as $n$ goes to infinity.

Note that similar analysis can be carried out for the two variants using $l_{2}-l_{0}$ regularization and approximate $l_{p}$ regularization mentioned in Section 4.2. 

When the overall cost function takes the form
\begin{equation} \label{eqwcf}
J(\bm{f},\bm{\phi})=\|\bm{g}-\bm{C}(\bm{\phi})\bm{f}\|_{2}^{2}+\lambda\sum_{i=1}^{N}(1-e^{-\frac{|\bm{f}_{i}|^{2}}{2\delta^{2}}}),
\end{equation}
the corresponding $K(\bm{b},\bm{f},\bm{\phi})$ is constructed as:
\begin{equation} \label{eqwK}
K(\bm{b},\bm{f},\bm{\phi})=\|\bm{g}-\bm{C}(\bm{\phi})\bm{f}\|_{2}^{2}+\lambda\sum_{i=1}^{N}[(|\bm{f}_{i}|^{2}-2\delta^{2})\bm{b}_{i}+2\delta^{2}\bm{b}_{i}\ln{(2\delta^{2}\bm{b}_{i})}+1].
\end{equation}
And this leads to:
\begin{equation} \label{eqwbi}
\bm{b}_{i}^{*}=\frac{e^{-\frac{|\bm{f}_{i}|^{2}}{2\delta^{2}}}}{2\delta^{2}}.
\end{equation}

When the overall cost function takes the form
\begin{equation} \label{eqgmcf}
J(\bm{f},\bm{\phi})=\|\bm{g}-\bm{C}(\bm{\phi})\bm{f}\|_{2}^{2}+\lambda\sum_{i=1}^{N}\frac{|\bm{f}_{i}|^{2}}{|\bm{f}_{i}|^{2}+\delta^{2}},
\end{equation}
the corresponding $K(\bm{b},\bm{f},\bm{\phi})$ is can then be constructed as:
\begin{equation} \label{eqgmK}
K(\bm{b},\bm{f},\bm{\phi})=\|\bm{g}-\bm{C}(\bm{\phi})\bm{f}\|_{2}^{2}+\lambda\sum_{i=1}^{N}[(|\bm{f}_{i}|^{2}+2\delta^{2})\bm{b}_{i}-2\sqrt{2}\delta \sqrt{\bm{b}_{i}}+1].
\end{equation}
And this leads to:
\begin{equation} \label{eqgmbi}
\bm{b}_{i}^{*}=\frac{2\delta^{2}}{(|\bm{f}_{i}|^{2}+2\delta^{2})^{2}}.
\end{equation}

Whereas when the overall cost function takes the form
\begin{equation} \label{eqaplcf}
J(\bm{f},\bm{\phi})=\|\bm{g}-\bm{C}(\bm{\phi})\bm{f}\|_{2}^{2}+\lambda\sum_{i=1}^{N}(|\bm{f}_{i}|^{2}+\beta)^{\frac{p}{2}},
\end{equation}
the corresponding $K(\bm{b},\bm{f},\bm{\phi})$ can be constructed as:
\begin{equation} \label{eqaplK}
K(\bm{b},\bm{f},\bm{\phi})=\|\bm{g}-\bm{C}(\bm{\phi})\bm{f}\|_{2}^{2}+\lambda\sum_{i=1}^{N}[\bm{b}_{i}(|\bm{f}_{i}|^{2}+\beta)+\frac{2-p}{2}(\frac{2\bm{b}_{i}}{p})^{\frac{p}{p-2}}].
\end{equation}
And this leads to:
\begin{equation} \label{eqaplbi}
\bm{b}_{i}^{*}=\frac{p}{2(|\bm{f}_{i}|^{2}+\beta)^{1-\frac{p}{2}}}.
\end{equation}

\section{Experimental results}
\label{sec:typestyle}

For the numerical experiments in this paper, the same radar system model as in \cite{onhon2011sparsity} is used, whose parameters are listed in Table \ref{tab:radarparameters}.

\begin{table}[!htbp]
	\scriptsize
	\caption{\textbf{Parameters of the radar system.}}
	\begin{center}
		\begin{tabular}{|l|c|c|}
			\hline
			\textbf{Carrier Frequency}&$2\pi \times 10^{10} rad/s$\\\hline
			\textbf{Chirp Rate}&$2\pi \times 10^{12} rad/s^{2}$\\\hline
			\textbf{Pulse Duration}&$4 \times 10^{-4}s$\\\hline
			\textbf{Angular Range}&$\ang{2.3}$\\\hline
		\end{tabular}
	\end{center}
	\label{tab:radarparameters}
\end{table}

In each experiment, this radar system model is used to generate a simulated phase history from a given reflectivity scene. This phase history is then corrupted by adding 1D random phase error along the azimuth direction as well as white Gaussian noise to it. This corrupted phase history is used to reconstruct a SAR image while correcting for the phase error.

\begin{figure}[t]
	\renewcommand{\figurename}{\textbf{Fig.}}
	\centering
	\subfigure[]{
		\includegraphics[width=1in]{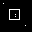}}
	\subfigure[]{
		\includegraphics[width=1in]{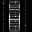}}
	\subfigure[]{
		\includegraphics[width=1in]{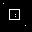}}
	\subfigure[]{
		\includegraphics[width=1in]{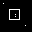}}
	\subfigure[]{
		\includegraphics[width=1in]{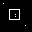}}
	\captionsetup{font={scriptsize}}
	\caption{Visual results for Scene 1, a simulated $32\times 32$ scene obtained by various methods. (a) original scene, (b) polar format reconstruction, (c) SDA, (d) WAMA, (e) CFBA.}
\end{figure} 

\begin{figure}[t]
	\renewcommand{\figurename}{\textbf{Fig.}}
	\centering
	\subfigure[]{
		\includegraphics[width=1in]{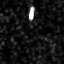}}
	\subfigure[]{
		\includegraphics[width=1in]{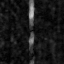}}
	\subfigure[]{
		\includegraphics[width=1in]{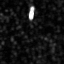}}
	\subfigure[]{
		\includegraphics[width=1in]{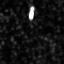}}
	\subfigure[]{
		\includegraphics[width=1in]{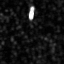}}
	\captionsetup{font={small}}
	\\
	\subfigure[]{
		\includegraphics[width=1in]{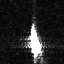}}
	\subfigure[]{
		\includegraphics[width=1in]{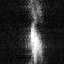}}
	\subfigure[]{
		\includegraphics[width=1in]{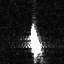}}
	\subfigure[]{
		\includegraphics[width=1in]{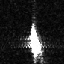}}
	\subfigure[]{
		\includegraphics[width=1in]{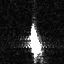}}
	\captionsetup{font={scriptsize}}
	\caption{Visual results for Scene 2 (a $64\times 64$ patch from TerraSAR-X) and Scene 3 (a $64\times 64$ patch from Sentinel-1) obtained by various methods. First row for Scene 2, the second row for Scene 3. Each row from left to right: original scene, polar format reconstruction, SDA, WAMA, CFBA.}
\end{figure} 

\begin{figure}[t]
	\renewcommand{\figurename}{\textbf{Fig.}}
	\centering
	\subfigure[]{
		\includegraphics[width=1in]{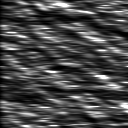}}
	\subfigure[]{
		\includegraphics[width=1in]{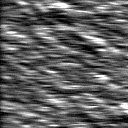}}
	\subfigure[]{
		\includegraphics[width=1in]{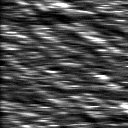}}
	\subfigure[]{
		\includegraphics[width=1in]{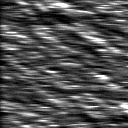}}
	\subfigure[]{
		\includegraphics[width=1in]{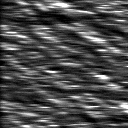}}
    \\
	\subfigure[]{
		\includegraphics[width=1in]{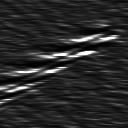}}
	\subfigure[]{
		\includegraphics[width=1in]{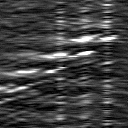}}
	\subfigure[]{
		\includegraphics[width=1in]{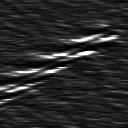}}
	\subfigure[]{
		\includegraphics[width=1in]{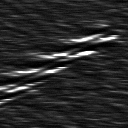}}
	\subfigure[]{
		\includegraphics[width=1in]{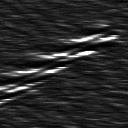}}
	\captionsetup{font={scriptsize}}
	\caption{Visual results for Scene 4 and Scene 5, two $128\times 128$ simulated sea surfaces, obtained by various methods. First row for Scene 4, the second row for Scene 5. Each row from left to right: original scene, polar format reconstruction, SDA, WAMA, CFBA.}
\end{figure} 

We compare the performance of four methods in each experiment. The first method is the traditional polar format algorithm \cite{walker1980range} which doesn't involve a process of autofocusing, and is therefore expected to result in a blurry formed image as a result of the phase error added into the simulated phase history. The second method is the sparsity driven autofocus (SDA) method of \cite{onhon2011sparsity} (we choose approximate $l_{1}$ norm as the regularizer of their cost function as an example), a state of the art SAR autofocusing technique operating in an inverse problem framework similar to the one proposed in this paper. The remaining two methods are the WAMA method in \cite{zhang2021sar} and the proposed CFBA method. 

Apart from visual comparison, two numerical metrics are also computed to better assess the performance of each method. One is the mean square error (MSE) between the reconstructed SAR image (using the corrupted phase history) and the ground truth (the reconstructed SAR image from the un-corrupted phase history). The second metric we employ is the entropy of the reconstructed SAR image, as an indicator of sharpness. For both of these two metrics, smaller values indicate better performance. For all the compared methods with tunable parameters, we present the result corresponding to the setting of the parameters which gives the best MSE value for that method.

In the first experiment, we use a simulated scene measuring $32\times 32$ pixels. The visual results are presented in Fig.1, and the numerical results are listed in Table 2. It can be observed that the visual result of polar format suffers from severe defocusing effect, making it impossible to discern the targets. However, the reconstructed images by SDA, WAMA, and the proposed CFBA method are all very focused and highly resemble the original scene. Since the values of MSE and entropy for WAMA and the proposed CFBA method are lower than SDA, their results are suggested to be sharper and more similar to the original scene.

In the second and the third experiment, a real SAR image from TerraSAR-X and another real SAR image from Sentinel-1 are used in place of the original scene. For both cases, due to the high computational burden of our method for scenes of large size (mainly due to the need to contrsuct the observation matrix $bm{C}$), a $64\times 64$ patch is cut from the original SAR image and regarded as an input scene. The corrupted pseudo-phase history is generated from it as described above. 

Fig. 2 shows the reconstructed images by all 4 methods for the second experiment (in the first row) and the third experiment (in the second row). Table 2 again contains all the corresponding results of the two numerical indices for these scenes. According to Fig. 2 and Fig. 3, the polar format algorithm once again gives reconstructed results with seriously smeared targets, especially notable in Fig. 2. In contract, SDA, WAMA and the proposed CFBA method can remove phase errors effectively and present focused targets, displaying significant improvement over the result of the polar format algorithm. Nevertheless, the results of the numerical indices in Table 2 demonstrate that WAMA and the proposed CFBA method both outperform SDA. 

In the fourth and the fifth experiment, two simulated images of the sea surface are used as the original scene, the first one only including sea waves, the second including a travelling ship and its wake as well. Simulated as the scenes are, they are not as simple as Scene 1, which is a mere combination of black and white regions resembling point reflectors, but are rather based on an exquisite model taking the most important SAR imaging effects into account \cite{rizaev2021modeling}. The scenes are based on a model of the sea surface using the Pierson-Moskowitz spectrum and cosine-squared spreading function with wind speed $V_{w} = 8$ m/s for the first image and $V_{w} = 4$ m/s for the second image, with waves traveling at $45^{\circ}$ relative to the SAR flight direction. For the second image, the size of the ship is 55 m, with 8 m beam and 3 m draft, moving at a velocity of 8 m/s at $45^{\circ}$ relative to the SAR flight direction. The original size of both SAR images is $1\times 1$ km with a spatial resolution of 1.25 m, while SAR platform parameters are as follows: platform altitude is 2.5 km, platform velocity is 125 m/s and incidence angle is $\theta_r$ = $35^{\circ}$, and signal parameters are X-band (9.65 GHz) and VV polarization. Both scenes shown here are of $128\times 128$ pixels, patches from the original images due to heavy computational burden, and their corresponding corrupted pseudo-phase histories are generated from them in the same way as aforementioned.

The visual results for all 4 methods for the fourth experiment and the fifth experiment are shown in Fig. 3 in the first row and the second row respectively. Since the original pixel values in the images are rather small, for visual convenience, "imadjust" function in Matlab is used before depicting. For the fourth experiment, the results of SDA, WAMA, and CFBA are with better contrast, i. e., the bright regions in (c), (d), (e) of Fig. 3 are brighter than those in (b), and their dark regions are darker. For the fifth experiment, (h), (i), (j) are sharper than (g) and display much more concentrated ship wakes. Meanwhile, the numerical results in Table 2 also demonstrate that CFBA, SDA, and WAMA give comparable performance. While WAMA is the best in the value of entropy, CFBA is the best in the value of MSE. 

Fig. 4 displays how $J(\bm{f}^{(n)},\bm{\phi}^{(n)})$ changes with increasing $n$ until the stopping criterion is satisfied for both WAMA method and the proposed CFBA method, taking the first three experiments as examples. In each sub-figure, the vertical axis represents $J(\bm{f}^{(n)},\bm{\phi}^{(n)})$, the value of the cost function (\ref{oc}) computed for $\bm{f}^{(n)}$ and $\bm{\phi}^{(n)}$, while the horizontal axis represents the iterative numbers $n$ in the loop of the alternating minimization. It can be seen that in all three experiments, $J(\bm{f}^{(n)},\bm{\phi}^{(n)})$ decreases monotonically for both CFBA and WAMA. This is consistent with the conclusions of our convergence analysis, and gives an experimental validation for the convergence of CFBA and WAMA in a sense.

\begin{table}[!t]\small
	\scriptsize
	\caption{\textbf{Numerical evaluation of the experimental results for all the methods.}}
	\begin{center}
		\begin{tabular}{|l|c|c|c|c|c|}
			\hline
			\multicolumn{6}{|c|}{\textbf{MSE}} \\ \hline
			\textbf{Method}&\textbf{Scene 1}&\textbf{Scene 2}&\textbf{Scene 3}&\textbf{Scene 4}&\textbf{Scene 5}\\ \hline
			\textbf{SDA}&5.4310$\times 10^{-6}$ & 6.4964$\times 10^{-5}$ & 6.3576 $\times 10^{-5}$ & 1.3997 $\times 10^{-5}$ & 6.8909 $\times  10^{-6}$\\ \hline
			\textbf{WAMA}&1.2227$\times 10^{-6}$ & 6.3029$\times  10^{-5}$ & 5.3663$\times 10^{-5}$ & 2.2250$\times 10^{-5}$ & 7.8785$\times 10^{-6}$\\ \hline
			\textbf{CFBA}&1.1836$\times 10^{-6}$ &6.2940$\times  10^{-5}$ &5.4803$\times 10^{-5}$ & 1.3483$\times 10^{-5}$ & 6.5628$\times 10^{-6}$\\ \hline
		    \multicolumn{6}{|c|}{\textbf{Entropy}} \\ \hline
			\textbf{Method}&\textbf{Scene 1}&\textbf{Scene 2}&\textbf{Scene 3}&\textbf{Scene 4}&\textbf{Scene 5}\\ \hline
			\textbf{SDA}&1.4621 &5.4410 &5.6918 &4.5720 &4.2847\\ \hline
			\textbf{WAMA}&0.3327 &5.4333 &5.6641 &4.5230 &4.2782\\ \hline
			\textbf{CFBA}&0.3430 &5.4228 &5.6602 &4.5617 &4.2916\\ \hline
		\end{tabular}
	\end{center}
	\label{tab:allres}
\end{table}

\begin{figure}[!htbp]
	\renewcommand{\figurename}{\textbf{Fig.}}
	\centering
	\subfigure[]{
		\includegraphics[width=1.6in]{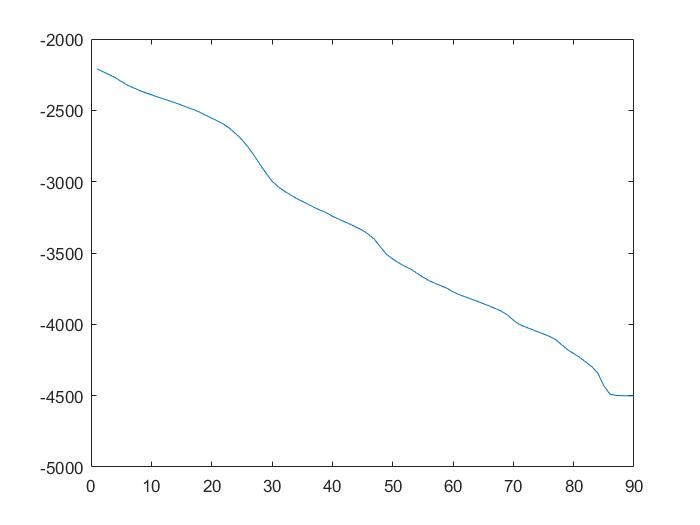}}
	\subfigure[]{
		\includegraphics[width=1.6in]{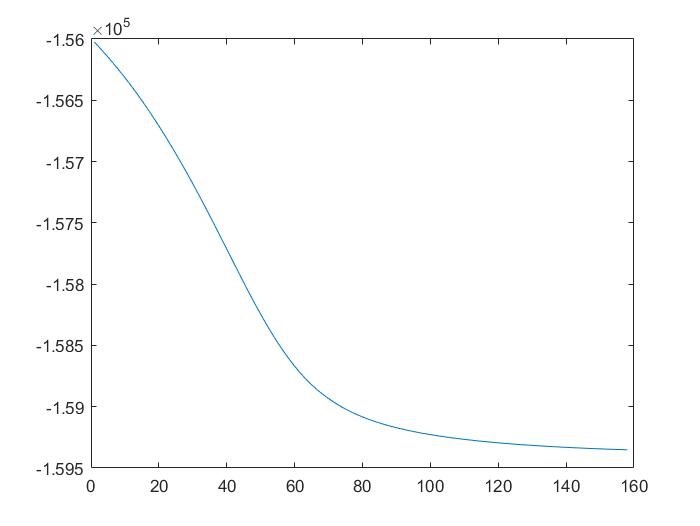}}
	\subfigure[]{
		\includegraphics[width=1.6in]{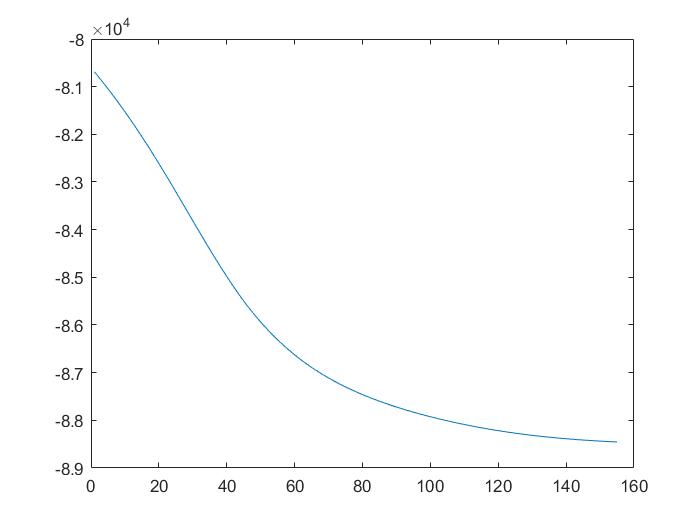}}
	\\
	\subfigure[]{
		\includegraphics[width=1.6in]{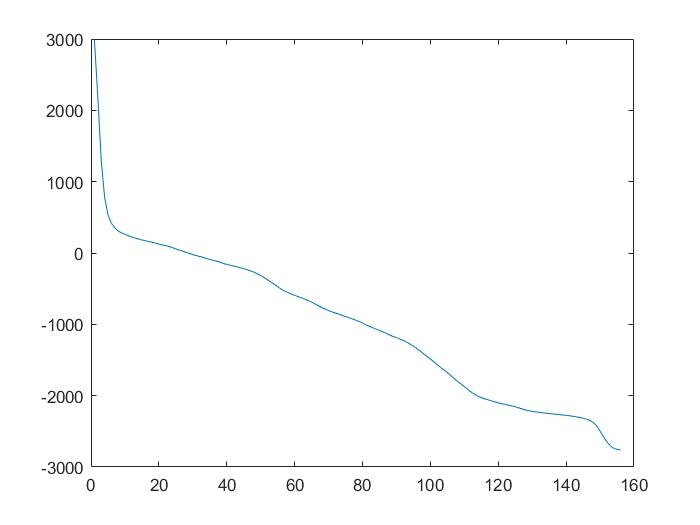}}
	\subfigure[]{
		\includegraphics[width=1.6in]{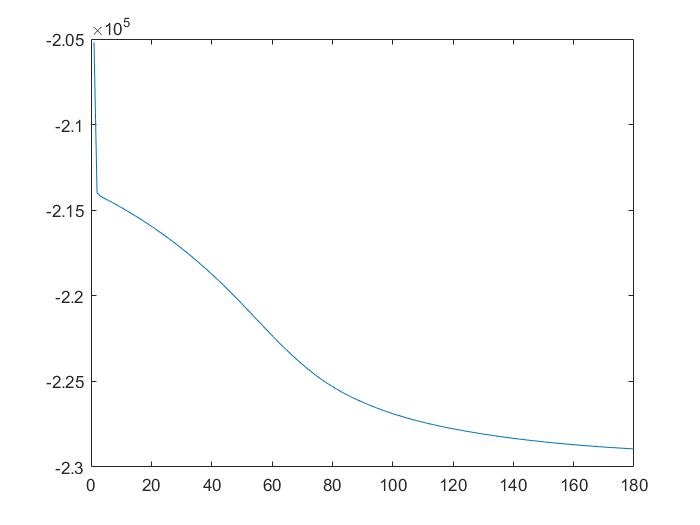}}
	\subfigure[]{
		\includegraphics[width=1.6in]{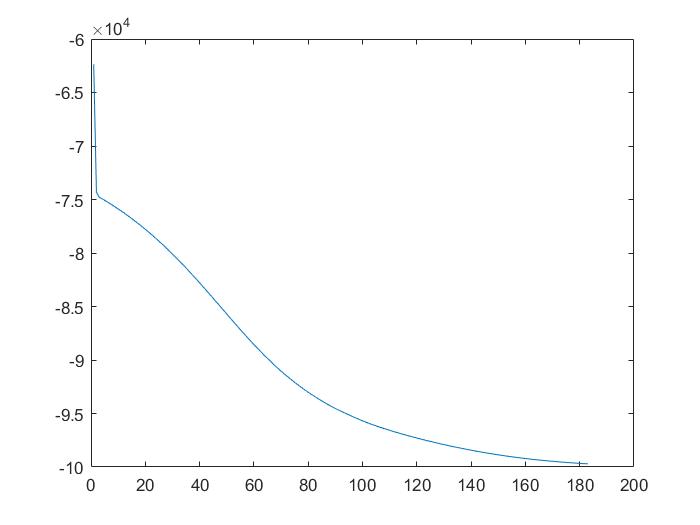}}
	\captionsetup{font={scriptsize}}
	\caption{The values of $J(f^{(n)},\phi^{(n)})$ until convergence. (a)-(c) CFBA for Scene 1, Scene 2, Scene 3; (d)-(f) WAMA for Scene 1, Scene 2, Scene 3.}
\end{figure}

\section{Conclusion}
In this paper, an optimization model regularized by magnitude Cauchy penalty is proposed to simultaneously reconstruct a SAR image and achieve autofocusing. An alternating minimization framework named CFBA is proposed to solve this inverse problem, in which the sub-problems related to the desired SAR image is solved by a complex forward-backward splitting method, and its convergence is analyzed. Besides, WAMA method based on Wirtinger calculus is reviewed and further discussed with regard to its extension and convergence. Experimental results on simulated phase histories derived from a simulated scene and several real SAR images demonstrate that the proposed CFBA method can reconstruct highly focused SAR images and effectively remove phase errors, showing performance competitive to WAMA.


%

\section{Appendix}

It has already been shown in the main text that $\widetilde{J}_{n}(\widetilde{\bm{f}})$ satisfies KL property. In order to apply this conclusion to $J_{n}(\bm{f})$, just notice that the definitions of $\nabla_{\bm{f}} J_{n}(\bm{f})$ and $\nabla \widetilde{J}_{n}(\widetilde{\bm{f}})$ imply $\|\nabla_{\bm{f}} J_{n}(\bm{f})\|=\frac{1}{2}\|\nabla \widetilde{J}_{n}(\widetilde{\bm{f}})\|$. 

Therefore, $J_{n}(\bm{f})$ is a KL function. It can also be shown that $J_{n}(\bm{f})=G(\bm{f})+H(\bm{f})$ is proper, continuous, and bounded from below; $H(\bm{f})$ is finite valued, differentiable, and has a Lipschitz continuous gradient; $G(\bm{f})$ is continuous on its domain. Apart from differentiability which is defined by Wirtinger calculus in a special way, the rest of these mentioned properties can be easily established in the complex case by directly replacing the real variable in the original definitions by a complex variable.

Now, we continue to show that all the three assumptions for Theorem 4.2 in \cite{attouch2013convergence} are satisfied. We point out that this is not a trivial task, because the inner product used in the original proof \cite{attouch2013convergence} takes only real values, which doesn't hold in our complex case. 

First, by computing the optimality condition of the Moreau envelope for $J_{n}(\bm{f})$, and let $\bm{v}^{(k+1)}\in \partial G(\bm{o}^{(k+1)})$, we have:
\begin{equation} \label{prc1}
\begin{split}
2\mu \bm{v}^{(k+1)}+2\mu \nabla H(\bm{o}^{(k)})+\bm{o}^{(k+1)}-\bm{o}^{(k)}=0,
\end{split}
\end{equation}
and therefore
\begin{equation} \label{prc2}
\begin{split}
\|\bm{v}^{(k+1)}+\nabla H(\bm{o}^{(k)})\|=\frac{1}{2\mu}\|\bm{o}^{(k+1)}-\bm{o}^{(k)}\|.
\end{split}
\end{equation}

For $H(\bm{f})=\|\bm{g}-\bm{C}^{(n)}\bm{f}\|_{2}^{2}$, according to the convexity of $H(\bm{f})$ \cite{zhang2015complex} and the property of $\nabla H(\bm{f})$ \cite{liu2020neurodynamic}, we have for any $\bm{f}_{1}$ and $\bm{f}_{2}$:
\begin{equation} \label{prc3}
H(\bm{f}_{1})-H(\bm{f}_{2})\le 2Re\{(\bm{f}_{1}-\bm{f}_{2})^{H}\nabla H(\bm{f}_{1})\}.
\end{equation}
Therefore,
\begin{equation} \label{prc4}
H(\bm{f}_{1})-H(\bm{f}_{2})-2Re\{(\bm{f}_{1}-\bm{f}_{2})^{H}\nabla H(\bm{f}_{2})\}\le 2Re\{(\bm{f}_{1}-\bm{f}_{2})^{H}(\nabla H(\bm{f}_{1})-\nabla H(\bm{f}_{2}))\}.
\end{equation}
Since $\nabla H(\bm{f})=(\bm{C}^{(n)})^{H}(\bm{C}^{(n)}\bm{f}-\bm{g})$, we have
\begin{equation} \label{prc5}
Re\{(\bm{f}_{1}-\bm{f}_{2})^{H}(\nabla H(\bm{f}_{1})-\nabla H(\bm{f}_{2})\}=\|\bm{C}^{(n)}(\bm{f}_{1}-\bm{f}_{2})\|_{2}^{2}.
\end{equation}
Therefore, the right side of (\ref{prc4}) is real, and thus
\begin{equation} \label{prc6}
\begin{split}
H(\bm{f}_{1})-H(\bm{f}_{2})-2Re\{(\bm{f}_{1}-\bm{f}_{2})^{H}\nabla H(\bm{f}_{2})\}\le 2(\bm{f}_{1}-\bm{f}_{2})^{H}(\nabla H(\bm{f}_{1})-\nabla H(\bm{f}_{2}))\\
\le 2\|(\bm{f}_{1}-\bm{f}_{2})\|\|\nabla H(\bm{f}_{1})-\nabla H(\bm{f}_{2}\| \le 2L\|\bm{f}_{1}-\bm{f}_{2}\|_{2}^{2}.
\end{split}
\end{equation}
As a result, we have
\begin{equation} \label{prc7}
\begin{split}
H(\bm{o}^{(k+1)})\le H(\bm{o}^{(k)})+2Re\{(\bm{o}^{(k+1)}-\bm{o}^{(k)})^{H}\nabla H(\bm{o}^{(k)})\}+2L\|\bm{o}^{(k+1)}-\bm{o}^{(k)}\|_{2}^{2}.
\end{split}
\end{equation}

On the other hand, from the definition of proximal operator,
\begin{equation} \label{prc8}
\mu G(\bm{o}^{(k+1)})+\frac{1}{2}\|\bm{o}^{(k+1)}-\bm{o}^{(k)}+2\mu \nabla H(\bm{o}^{(k)})\|_{2}^{2}\le \mu G(\bm{o}^{(k)})+\frac{1}{2}\|2\mu \nabla H(\bm{o}^{(k)})\|_{2}^{2}.
\end{equation}
Expanding (\ref{prc8}) yields
\begin{equation} \label{prc9}
\mu G(\bm{o}^{(k+1)})+\frac{1}{2}\|\bm{o}^{(k+1)}-\bm{o}^{(k)}\|_{2}^{2}+2\mu Re\{(\bm{o}^{(k+1)}-\bm{o}^{(k)})^{H}\nabla H(\bm{o}^{(k)})\}\le \mu G(\bm{o}^{(k)}),
\end{equation}
and therefore
\begin{equation} \label{prc10}
G(\bm{o}^{(k+1)})+\frac{1}{2\mu}\|\bm{o}^{(k+1)}-\bm{o}^{(k)}\|_{2}^{2}+ 2Re\{(\bm{o}^{(k+1)}-\bm{o}^{(k)})^{H}\nabla H(\bm{o}^{(k)})\}\le G(\bm{o}^{(k)}),
\end{equation}

Combine (\ref{prc7}) with (\ref{prc10}), we have
\begin{equation} \label{prc11}
\begin{split}
G(\bm{o}^{(k+1)})+H(\bm{o}^{(k+1)})+\frac{a-4L}{2}\|\bm{o}^{(k+1)}-\bm{o}^{(k)}\|_{2}^{2}\\
\le G(\bm{o}^{(k+1)})+H(\bm{o}^{(k)})+2Re\{(\bm{o}^{(k+1)}-\bm{o}^{(k)})^{H}\nabla H(\bm{o}^{(k)})\}+\frac{a}{2}\|\bm{o}^{(k+1)}-\bm{o}^{(k)}\|_{2}^{2}\\
\le G(\bm{o}^{(k)})+H(\bm{o}^{(k)}),
\end{split}
\end{equation}
with $a=\frac{1}{\mu}$. This is actually $J_{n}(\bm{o}^{(k+1)})+\frac{a-4L}{2}\|\bm{o}^{(k+1)}-\bm{o}^{(k)}\|_{2}^{2}\le J_{n}(\bm{o}^{(k)})$, as long as $a>4L$. In contrast, in the proof from real perspective the corresponding requirement is just $a>L$. Therefore, if some other appropriate techniques are utilized, it may be possible to get an inequality better than (\ref{prc7}) and obtain $a>L$.   

Second, using differential rule, we have $\bm{v}^{(k+1)}+\nabla h(\bm{o}^{(k+1)})\in \nabla J_{n}(\bm{o}^{(k+1)})$.

At last, with (\ref{prc2}), it can be deduced that
\begin{equation} \label{prc12}
\begin{split}
\|\bm{v}^{(k+1)}+\nabla H(\bm{o}^{(k+1)})\|
\le \|\bm{v}^{(k+1)}+\nabla H(\bm{o}^{(k)})\|+\|\nabla H(\bm{o}^{(k+1)})-\nabla H(\bm{o}^{(k)})\|\\
\le \frac{a}{2}\|\bm{o}^{(k+1)}-\bm{o}^{(k)}\|+L\|\bm{o}^{(k+1)}-\bm{o}^{(k)}\|.
\end{split}
\end{equation}

Now we are exactly in the case of Theorem 4.2 in \cite{attouch2013convergence} and the rest of the proof is similar (just formally substitute the real vectors therein by complex vectors). In conclusion, we can get the same result as Theorem 5.1 in \cite{attouch2013convergence}, i.e., the convergence of the sequence $\bm{o}^{(k)}$ to a critical point of $J_{n}(\bm{f})$.

Moreover, denote by $\bm{f}^{*}$ the global minimizer of $J_{n}(\bm{f})$. According to Theorem 2.12 in \cite{attouch2013convergence}, for each $r>0$, there exist $u\in (0,r), \delta>0$ such that the inequalities $\|\bm{o}^{(0)}-\bm{f}^{*}\|<u$ and $\min J_{n}(\bm{f})<J(\bm{o}^{(0)})<\delta+\min J_{n}(\bm{f})$ imply that the sequence $\bm{o}^{(k)}$ will converge to some $\bm{o}^{*}$ with $\bm{o}^{(k)}\in B(\bm{f}^{*}, r)$ for arbitrary $k$ and $J_{n}(\bm{o}^{*})=\min J_{n}(\bm{f})$.


\bibliographystyle{unsrt}  
\bibliography{refs}  






\end{document}